\documentclass[summary]{ursi}

\title{Robust Integrated Sensing and Communication Beamforming for Dual-functional Radar and Communications: Method and Insights \\ \textit{(Invited Paper)}}

\author{Ahmad Bazzi\affref{ref1}, Marwa Chafii\affref{ref1}\affref{ref2}}

\affiliation{%
  \aff{ref1}{Engineering Division, NYU Abu Dhabi, 129188, UAE; e-mail: ahmad.bazzi@nyu.edu}
  \aff{ref2}{NYU WIRELESS,
NYU Tandon School of Engineering, Brooklyn, 11201, NY, USA; e-mail: marwa.chafii@nyu.edu}
}

\usepackage{amsmath,amsfonts}

\usepackage{graphicx}
\usepackage{cite}
\usepackage{tikz}
\usepackage{xcolor}
\usepackage{pgfplots}
\usepackage{standalone}
\usepackage{xcolor}
\usepackage{hyperref}
\definecolor{mycolor1}{rgb}{0.90471,0.19176,0.19882}%
\definecolor{mycolor2}{rgb}{0.29412,0.54471,0.74941}%
\definecolor{mycolor3}{rgb}{0.37176,0.71765,0.36118}%
\definecolor{mycolor4}{rgb}{1.00000,0.54824,0.10000}%
\definecolor{mycolor5}{rgb}{0.86500,0.81100,0.43300}%
\definecolor{mycolor6}{rgb}{0.68588,0.40353,0.24118}%
\definecolor{mycolor7}{rgb}{0.97176,0.55529,0.77412}%
\definecolor{mycolor8}{rgb}{0.63647,0.37529,0.67529}%
\usetikzlibrary{shapes.geometric}
\usetikzlibrary{arrows.meta}
\pgfplotsset{compat=1.5}
\DeclareMathOperator{\Real}{Re}
\DeclareMathOperator{\opt}{opt}

\DeclareMathOperator{\rank}{rank}
\DeclareMathOperator{\SNR}{SNR}

\DeclareMathOperator{\bpsph}{bits/sec/Hz}
\DeclareMathOperator{\bpsphpu}{bits/sec/Hz/user}
\DeclareMathOperator{\dB}{dB}
\DeclareMathOperator{\Tr}{Tr}

\DeclareMathOperator{\vect}{vec}
\DeclareMathOperator{\SINR}{SINR}

\begin{document}

\maketitle

\begin{abstract}
This work presents a novel robust beamforming design dedicated for dual-functional radar and communication (DFRC) base stations (BSs) in the context of integrated sensing and communications (ISAC). The architecture is intended for circumstances with imperfect channel state information (CSI). Our suggested approach demonstrates several tradeoffs for joint radar-communication deployment. Due to the DFRC nature of the design, the beamformer can simultaneously point towards an intended target, while optimizing communication quality of service. We unveil several insights regarding closed form expressions, as well as optimality of the proposed beamformer. Lastly, simulation results demonstrate the effectiveness of the proposed ISAC beamformer.
\end{abstract}

\section{Introduction}


Integrated sensing and communications (ISAC) \cite{bazzi2,bazzi3} is among the features that will leave a distinct imprint on 6G, due to its dual capability of performing communications, as well as sensing the environment. 6G wireless systems are intended to provide a wide range of high-accuracy sensing services, including Wi-Fi sensing for smart homes, indoor localization for robot navigation, and radar sensing for self-driving automobiles\cite{6g-wireless-systems}. Indeed, ISAC cooperation and convergence is a critical issue that will open the door to hitherto unseen novel applications such as autonomous driving, robotics, and the unmanned aerial vehicle (UAV) industry. As a result, the coexistence of radar and communication systems has been a key focus of research. The interoperability of the radar and communication systems results in a variety of practical benefits, the most evident of which is spectrum efficiency.

This paper considers, for the first time, a practical setting where a dual-functional radar and communication (DFRC) base station (BS) aims at designing an ISAC beamformer at the BS with imperfect CSI. This will enable proper communication, while maximizing the transmit energy in a certain direction so as to perform parameter estimation (e.g. \cite{PE1,PE2,PE3,PE4,PE5,PE6}) on the received echo. We use the terminology \textit{ISAC beamformer} to emphasize that the design not only takes into account communication quality of services, but also radar ones. Indeed, robust beamforming (BFing) optimizations have been applied for traditional communication models, namely in the context of secrecy rate maximization for privacy related applications in a MISO setting, simultaneous wireless information and power transfer for RF energy harvesting, non-orthogonal multiple access (NOMA) cognitive radio systems, etc. The key essence of beamforming designs relies on the assumption that the BSs have access to perfect CSI of the different communication users. In practical frameworks, however, the BSs can never have full perfect CSI, due to, for example, imperfect channel estimation and quantization errors. To the best of our knowledge, imperfect CSI for ISAC beamforming, in the context of DFRC, have not yet been addressed.

In this study, we are interested in BFing design dedicated for DFRC MIMO systems, where the scene consists of a DFRC BS, multiple communication users, and an intended target. The following work assumes imperfect channel state information (CSI) at the DFRC BS, when applying BFing towards the communication users and performing target detection using the same transmit waveform. We propose an optimization framework allowing the DFRC BS to perform the aforementioned tasks. The initial problem involves stochastic constraints, which are then transformed to deterministic forms through a series of carefully chosen relaxations. Our contributions involve (i) DFRC beamforming design with imperfect CSI, where the precoded matrix is aware of imperfect CSI (ii) closed-form solutions of DFRC ISAC beamforming design, which facilitates solving the proposed optimization problem in an online way, and (iii) simulation results demonstrating the effectiveness of the proposed DFRC design. An extended version of this work containing more details and proofs, has been reported in \cite{bazzi-outage}.

The rest of this paper is organized as follows. Section 2 presents the ISAC system model for both radar and communication sub-systems. Section 3 derives the methods and highlights  various insights related to the proposed robust DFRC beamformer. Section 4 presents our simulation results. Section 5 concludes the paper.

\textbf{Notation}: Upper-case and lower-case boldface letters denote matrices and vectors, respectively. $(.)^T$, $(.)^*$ and $(.)^H$ represent the transpose, the conjugate and the transpose-conjugate operators. The statistical expectation is $\mathbb{E}\lbrace . \rbrace$. The vectorization and unvectorization operators are denoted as $\vect$ and $\vect^{-1}$, respectively. In particular, $\vect$ takes an $N\times M$ matrix $\pmb{X}$ as input and returns an $NM \times 1$ vector, by stacking the columns of $\pmb{X}$. For any complex number $z \in \mathbb{C}$, the magnitude is $\vert z \vert$, its angle is $ \angle z$, and its real part is $\Real(z)$. The Frobenius norm of matrix $\pmb{X}$ is $\Vert \pmb{X} \Vert$. We denote a positive semi-definite matrix as $\pmb{X} \succeq \pmb{0}$. The matrix $\pmb{I}_N$ is the identity matrix of size $N \times N$. The zero-vector is $\pmb{0}$. The $\rank$ returns the rank of a matrix and $\Tr$ returns the trace of a matrix. The inverse of a square matrix is $\pmb{X}^{-1}$. The probability of an event $\mathcal{A}$ is $\Pr(\mathcal{A})$.

\begin{figure}[htbp]
  \centering
  \includegraphics[width=82mm]{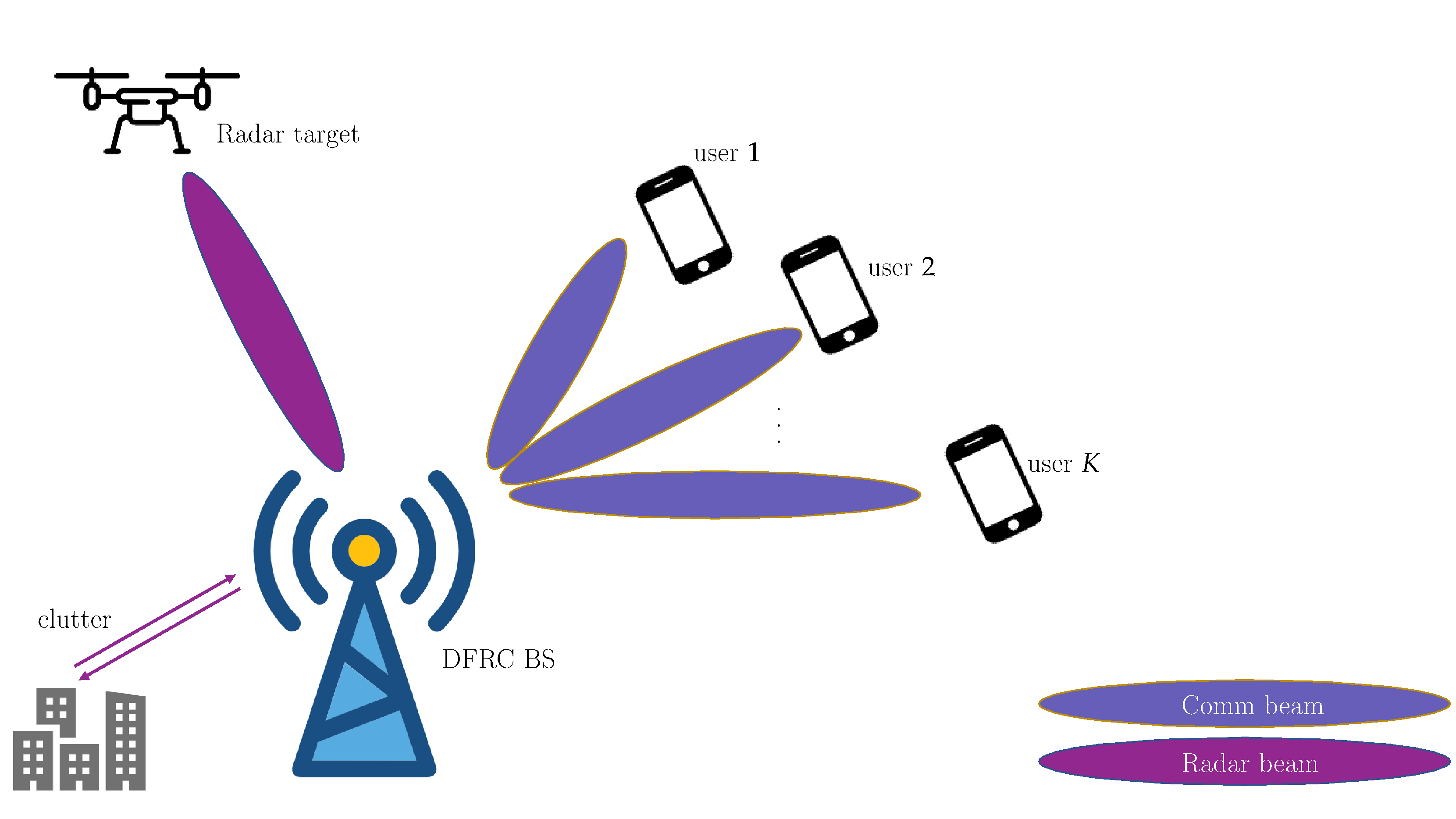}
  \caption{DFRC scenario including a DFRC base station, an intended target and $K$ communication users.}
  \label{fig_1}
\end{figure}

\section{System Model}

In this section, we introduce the communication and radar system models. To begin with, the communication model at the $n^{th}$ snapshot of the downlink is expressed as follows
\begin{equation}
	\label{eq:y=Hxpz}
	\pmb{y}_c[n] = \widetilde{\pmb{H}}\pmb{x}[n] + \pmb{z}_c[n] ,
\end{equation}
where  $\pmb{y}_c[n] \in \mathbb{C}^{K \times 1}$ is the vector of received signals over all communication users. The channel matrix is given by 
\begin{equation}
\widetilde{\pmb{H}} = \begin{bmatrix} \tilde{\pmb{h}}_1 & \tilde{\pmb{h}}_2 & \ldots & \tilde{\pmb{h}}_K \end{bmatrix}^T \in \mathbb{C}^{K \times N},	
\end{equation}
where $\widetilde{\pmb{H}}$ is flat Rayleigh type fading, and $\pmb{z}_c[n]$ is AWGN noise with noise variance $\sigma_c^2$. Furthermore, the transmit signal vector is denoted by $\pmb{x}[n]$. To model imperfect CSI, we follow \cite{wang2014outage}, such that $\tilde{\pmb{h}}_k = {\pmb{h}}_k + \Delta {\pmb{h}}_k$, where $\tilde{\pmb{h}}_k $ is the actual channel vector, $ {\pmb{h}}_k $ is the one assumed at the DFRC and $\Delta {\pmb{h}}_k$ is the channel estimation error assumed to be Gaussian with zero mean and covariance $\sigma_{\Delta \pmb{h}_k}^2 \pmb{I}$. On the other hand, the radar model at the $n^{th}$ snapshot reads
\begin{equation}
	\label{eq:yr}
	\pmb{y}_r[n] = \gamma_0 \pmb{a}(\theta_0)\pmb{a}^T(\theta_0)\pmb{x}[n] + \pmb{z}_r[n],
\end{equation}
where $y_r[n]$ is the received echo received at the angle-of-arrival $\theta_0$ and $\gamma_0$ is the reflection coefficient. The beamforming matrix $\pmb{W}$ to be designed is precoded as $\pmb{x}[n] = \pmb{W}\pmb{s}[n]$, with $\pmb{s}[n]$ being the constellation symbols. Also, the vector $\pmb{a}(\theta)$ is the steering vector, i.e. the array response to an angle $\theta$. For instance, in uniform linear array (ULA) configurations, the steering vector is given as 
	\begin{equation}
	    \pmb{a}(\theta) 
	    =
	    \begin{bmatrix}
	        1  & e^{j \frac{2\pi}{\lambda} d \sin(\theta)} & \ldots & e^{j \frac{2\pi}{\lambda} (N-1) d \sin(\theta)}
	    \end{bmatrix}^T,
	\end{equation}
	where $\lambda$ is the wavelength of the signal and $d$ is the inter-element spacing.


\begin{figure}[htb]
	 \centering
  \includegraphics[width=82mm]{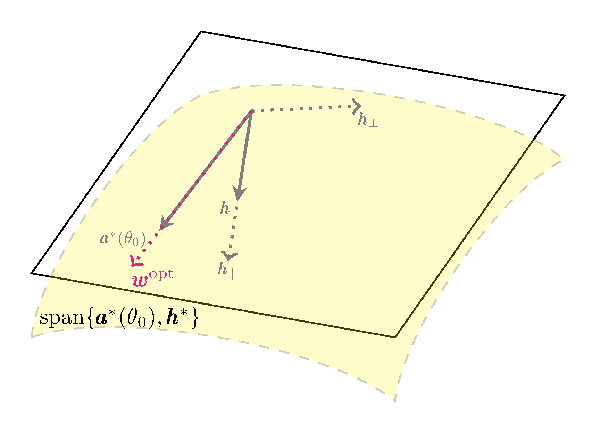}
  \includegraphics[width=82mm]{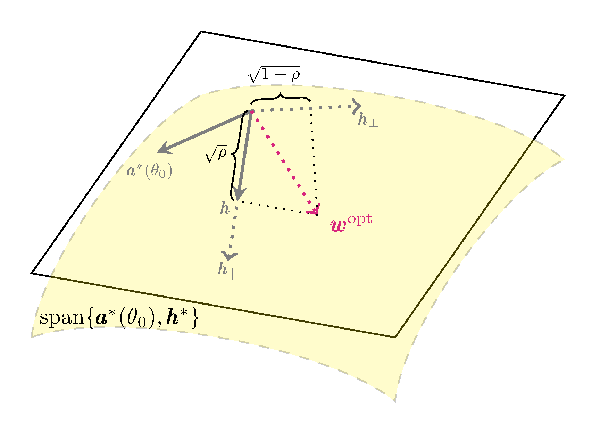}
  \caption{Graphical illustration of the DFRC ISAC beamforming weight vector, as a function of steering vector and channel response. Top: Case where $\Lambda \leq \vert \pmb{h}^T \pmb{a}^*(\theta_0) \vert^2$ and Bottom: Case where $\Lambda > \vert \pmb{h}^T \pmb{a}^*(\theta_0) \vert^2$.}
  \label{fig:subspace}
\end{figure}

\section{Proposed Robust DFRC Beamformer}
The proposed integrated sensing and communication problem, where the main cost function aims at maximizing the output of a Bartlett beamformer subject to a series of QoS constraints on all communication users involved in the system, is the following:
\begin{equation}
 \label{eq:problem1}
\begin{aligned}
(\mathcal{P}):
\begin{cases}
\max\limits_{\lbrace \pmb{w}_k \rbrace}&   P(\theta_0)\\
\textrm{s.t.}
 &  \Pr( \SINR_k \leq \gamma_k ) \leq p_k , \quad \forall k \\
 & \sum\limits_{k=1}^K \Tr(\pmb{W}_k) \leq 1, \\ 
 & \pmb{W}_k = \pmb{w}_k\pmb{w}_k^H  , \quad \forall k , \\
\end{cases}
\end{aligned}
\end{equation}
where $P(\theta) = \sum\limits_{k=1}^K \pmb{a}^T(\theta) \pmb{W}_k \pmb{a}^*(\theta)$ is the output of the Bartlett (sum-and-delay) beamforming towards angle $\theta$. Intuitively speaking, we aim at maximizing the output power of the Bartlett beamformer under the following constraints: The first constraint reflects per-user $\SINR$ outage to guarantee an acceptable rate, i.e. $\gamma_k$, under a probability of failure $p_k$. Furthermore, the second constraint is a power budget constraint on the transmit beamformer. Finally, the last constraint is to ensure a $\rank$-1 solution of the beamforming matrices. To this end, the above problem is non-convex due to the rank-1 constraint and untractable due to the probabilistic constraints. After many relaxations and approximations (see details in \cite{bazzi-outage}), we get
\begin{equation}
\label{eq:problem1-linear-proba-bernstien-socp-relax-final}
\begin{aligned}
\begin{cases}
\max\limits_{\Big\lbrace\substack{ \pmb{W}_k \\ \nu_k \\ \mu_k }\Big\rbrace}&   P(\theta_0)\\
\textrm{s.t.}
 & 
\sigma_k^2 \leq \Tr(\pmb{A}_k) - \sqrt{2\epsilon_k}\mu_k - \epsilon_k \nu_k, \quad \forall k  	 \\
   & \sigma_k^2 = \sigma_c^2   -  {\pmb{h}}_k^T \bar{\pmb{W}}_{k} {\pmb{h}}_k^* , \\
   & \nu_k \geq 0, \quad \forall k \\
    & \nu_k  \pmb{I}_N  \succeq - \pmb{A}_k, \quad \forall k  \\
    & \hspace{-0.25cm}  \left[\begin{array}{c|c}
		\mu_k  & 
	\begin{array}{cc}
		\sqrt{2}\pmb{b}_k^H &
		\vect(\pmb{A}_k)^H
	\end{array} \\
	\hline
	\begin{array}{cc}
		\sqrt{2}\pmb{b}_k \\
		\vect(\pmb{A}_k)
	\end{array} &
	\mu_k \pmb{I}_{N+N^2}
	\end{array}
	\right]\succeq
	\pmb{0} \\ 
    & \pmb{W}_k \succeq \pmb{0}, \quad \forall k, \sum\limits_{k=1}^K \Tr(\pmb{W}_k) \leq 1,
\end{cases}
\end{aligned}
\end{equation}
where $\pmb{A}_k = \sigma_{\Delta \pmb{h}_k}^2 \bar{\pmb{W}}_{k}$, $\pmb{b}_k = \sigma_{\Delta \pmb{h}_k} \bar{\pmb{W}}_{k}{\pmb{h}}_k^*$, $\bar{\pmb{W}}_{k}  = \frac{1}{\gamma_k} \pmb{W}_k - \sum\limits_{\ell = 1, \ell \neq k}^K \pmb{W}_k$ and $\epsilon_k = -\log(p_k)$.

Notice that the above problem in equation \eqref{eq:problem1-linear-proba-bernstien-socp-relax-final} is a convex optimization one, which can be resolved by classical convex optimization solvers (e.g. CVX and Gurobi). Now, we have the following theorems related to the problem at hand, i.e. the $\rank$-1 property of the relaxed problem

\textbf{Theorem (Rank-one Optimality): } \textit{Consider the convex optimization problem given in equation \eqref{eq:problem1-linear-proba-bernstien-socp-relax-final}. Then, for all $k = 1 \ldots K$, the solution of the problem in \eqref{eq:problem1-linear-proba-bernstien-socp-relax-final} satisfies }
\begin{equation}
	\rank(\pmb{W}_k^{\opt}) = 1 .
\end{equation}
\textbf{Proof}: See \cite{bazzi-outage}

The above theorem tells us that \textit{even though we had dropped the $\rank-$1 constraint in the problem defined in equation \eqref{eq:problem1-linear-proba-bernstien-socp-relax-final}, we are still guaranteed to obtain a $\rank-$1 optimal solution}. Meanwhile, we also derive closed form single user (SU) expressions as follows\footnote{We have omitted subscripts in the theorem due to the SU assumption}:

\textbf{Theorem (Single-User BFing in Closed-Form):} \textit{Let $K = 1$, the optimal solution to problem $(\mathcal{P}_6^{\text{SU}})$, is given as follows}
\begin{equation}
	\pmb{w}^{\opt}
	=
	\begin{cases}
		\frac{\pmb{a}^*(\theta_0)}{\Vert \pmb{a}(\theta_0) \Vert},   \qquad\text{if } \Lambda \leq     
 \vert {\pmb{h}}^T \pmb{a}^*(\theta_0) \vert^2 \\
	\sqrt{\rho}\frac{\pmb{h}_{\parallel}^H \pmb{a}^*(\theta_0)}{\Vert \pmb{h}_{\parallel}^H \pmb{a}^*(\theta_0) \Vert} \pmb{h}_{\parallel} + 
	\sqrt{1-\rho}\frac{\pmb{h}_{\perp}^H \pmb{a}^*(\theta_0)}{\Vert \pmb{h}_{\perp}	^H \pmb{a}^*(\theta_0) \Vert}\pmb{h}_{\perp} , \qquad   \text{else}
	\end{cases}
\end{equation}
\textit{where} $\Lambda = N(\gamma\sigma_c^2   - \sigma_{\Delta}^2 + \sqrt{2\epsilon}\sigma_{\Delta}\sqrt{\sigma_{\Delta}^2 + 2 \frac{\vert {\pmb{h}}^T \pmb{a}^*(\theta_0) \vert^2}{N}} )$. \textit{Furthermore, $\pmb{h}_{\parallel}$ is a normalized vector pointing towards $\pmb{h}$ and $\pmb{h}_{\perp}$ is normalized and orthogonal to $\pmb{h}_{\parallel}$.}
\textit{Moreover, $\sqrt{\rho}$ is a root of the following equation in the interval $[0,1]$}
\begin{equation}
	g(x) = x^2 \Vert \pmb{h} \Vert^2 - (\gamma \sigma_c^2 - \sigma_{\Delta}^2)-\sigma_{\Delta}\sqrt{2\epsilon}\sqrt{\sigma_{\Delta}^2 + x^2 \Vert \pmb{h} \Vert^2}, 
\end{equation}
\textit{where} $\sigma_{\Delta} = \sigma_{\Delta \pmb{h}}$. \\\\
\textbf{Proof}: See \cite{bazzi-outage}

The above theorem is particularly important in characterizing the trade-off between the radar and communication performances. Specifically, when the correlation between the channel and the steering vector towards the intended target, i.e. $\vert {\pmb{h}}^T \pmb{a}^*(\theta_0) \vert^2$, is above a certain threshold, then the solution is a purely radar one, i.e. $\frac{\pmb{a}^*(\theta_0)}{\Vert \pmb{a}(\theta_0) \Vert}$. This is intuitive as when $h$ and $\pmb{a}(\theta_0)$ have close resemblance, where the degree of similarity is decided by $\Lambda$, then the optimization problem focuses on maximizing the radar cost. On the other hand, when the correlation between the channel and the steering vector towards the intended target, i.e. $\vert {\pmb{h}}^T \pmb{a}^*(\theta_0) \vert^2$, is below $\Lambda$, then the solution lives in the subspace spanned by $\lbrace \pmb{h}_{\parallel}, \pmb{h}_{\perp} \rbrace$. Also, $\rho$ plays the role of defining the strength of $\pmb{w}^{\opt}$ within directions $\pmb{h}_{\parallel}$ or $\pmb{h}_{\perp}$. We illustrate this in Fig. \ref{fig:subspace}.

\section{Simulation Results}
\begin{figure}[htbp]
  \centering
  \includegraphics[width=82mm]{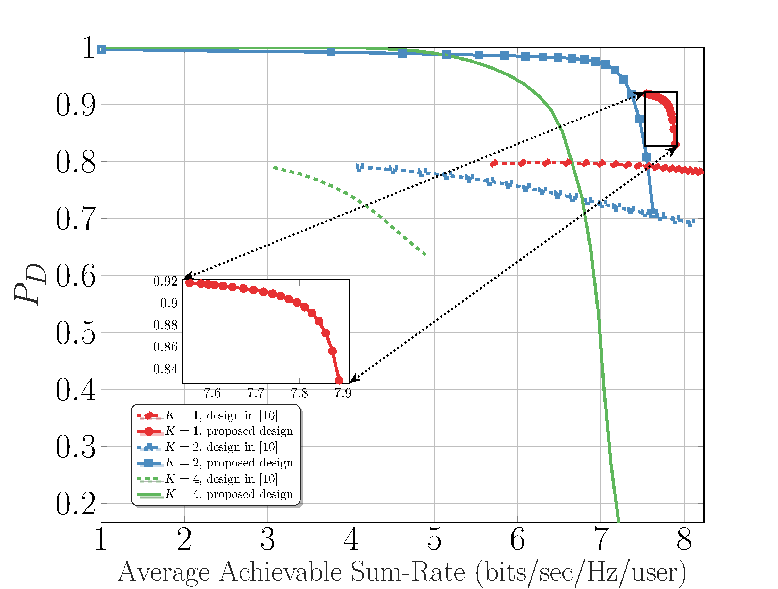}
  \caption{Trade-off between the average achievable sum rate per user and the radar detection
probability for $N = 5$ at receive $\SNR_r = 1 \dB$, and $P_{\text{FA}} = 10^{-4}$.}
  \label{fig:PD-rate-tradeoff}
\end{figure}

\begin{figure}[htbp]
  \centering
  \includegraphics[width=92mm]{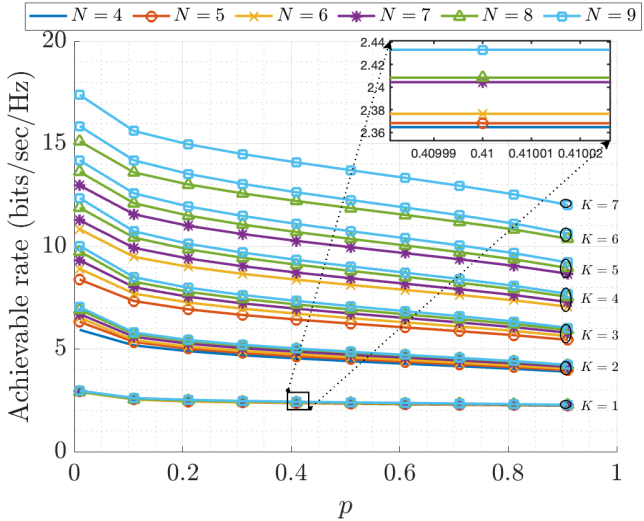}
  \caption{Achievable rates as a function of $p$ for different values of $N$ and $K$.}
  \label{fig:rate-vs-p}
\end{figure}

A tradeoff we can study is the probability of detection $P_D$, vs the average achievable sum rate.  The radar $\SNR$ is defined as $\SNR_r$ and is fixed to $1\dB$. We set $N = 5$ antennas. This is demonstrated in Fig. \ref{fig:PD-rate-tradeoff}. To show the superiority of our design, we compare to the beamforming design in \cite{toward-dfrc}. Referring to Fig. \ref{fig:PD-rate-tradeoff}, it is clear that a trade-off exists between the detection performance of the radar and the communication rate. In other words, for a fixed $P_D$, the rate increases with a decrease in number of users. Second, it should be noted that our design presents better tradeoffs in both, probability of detection and average achievable rate, especially when $K$ grows large. For example, with $K=4$ and an average achievable sum-rate of $4.78 \bpsphpu$, we have that $P_D \simeq 0.65$, when the design in \cite{toward-dfrc} is adopted; compared to $P_D \simeq 0.99$ using the proposed design, herein. 

In Fig. \ref{fig:rate-vs-p}, we aim at studying the impact of $p$ on the total achievable rate with different number of transmit antennas and communication users. It is clear that increasing $p$ leads to a degradation in the total achievable rate. In particular, a breakout region is observed around $p = 0.1$, where beyond this value, a stable slope contributes in the decrease of the total achievable rate. On another note, the degrees of freedom (DoF) available accounts for additional improvement in terms of $\SINR$, due to additional nulling of the interference terms; therefore, an increase in the total achievable rate. As an example, for the case of $K=2$, doubling the number of antennas contribute to an increase of roughly $0.3\bpsph$, whereas the same rate increase could be attained by adding only $1$ antenna, when $K = 5$.

\section{Conclusions}
In this paper, we have proposed a robust ISAC beamformer well-suited for imperfect channel state information situations. The beamformer can achieve high data rates, while simultaneously detecting passive targets in the scene. The paper also reveals different insights around the nature of the solution of the proposed ISAC beamformer in certain cases. Moreover, simulation results demonstrate the effectiveness of the proposed ISAC beamformer, when compared to state-of-the-art ones.

%
%
%
%

\end{document}